\documentclass[universe,article,accept,moreauthors,pdftex]{mdpi}
\firstpage{1} 
\pubvolume{1}
\issuenum{1}
\articlenumber{0}
\pubyear{2021}
\copyrightyear{2021}
\externaleditor{{Academic Editor}: Nicolas Chamel}
\datereceived{30 September 2021}
\dateaccepted{18 November 2021}
\datepublished{}
\hreflink{https://doi.org/} 

\usepackage{graphicx} 
\usepackage{subcaption}
\usepackage{datatool}

\DTLloaddb{acronyms}{test.csv}

\DTLsort{Acronym}{acronyms}
\Title{The Effect of Charge, Isospin, and Strangeness in the QCD Phase Diagram Critical End Point}

\TitleCitation{The Effect of Charge, Isospin, and Strangeness in the QCD Phase Diagram Critical End Point}

\Author{{Krishna Arya}l $^{1}$, Constantinos Constantinou $^{2,3}$, Ricardo L. S. Farias $^{4}$ and Veronica Dexheimer $^{1,*}$} 
\AuthorCitation{Aryal, K.; Constantinou, C.; Farias, R.L.S.; Dexheimer, V.}

\address{\\
$^{1}$ \quad Department of Physics, Kent State University, Kent, OH 44242, USA; karyal@kent.edu\\
$^{2}$ \quad INFN-TIFPA, Trento Institute of Fundamental Physics and Applications, 38123 Povo, TN, Italy;  cconstantinou@ectstar.eu\\ 
$^{3}$ \quad European Centre for Theoretical Studies in Nuclear Physics and Related Areas, 38123 Villazzano, TN, Italy \\
$^{4}$ \quad Departamento de F\'{i}sica, Universidade Federal de Santa Maria, Santa Maria, RS 97105-900, Brazil; ricardo.farias@ufsm.br
}

\corres{Correspondence: vdexheim@kent.edu}

\abstract{In this work, we discuss the  deconfinement phase transition to quark matter in hot/dense matter. We {examine} the effect that different charge fractions, isospin fractions, net strangeness, and chemical equilibrium with respect to leptons have on the position of the coexistence line between different phases. In particular, we investigate how different sets of conditions that describe matter in neutron stars and their mergers, or matter created in heavy-ion collisions affect the position of the critical end point, namely where the first-order phase transition becomes a crossover. We also present an introduction to the topic of critical points, including a review of recent {advances} concerning QCD critical points.}

\begin{document}
\begin{filecontents*}{test.csv}
Acronym, Description
QCD,Quantum Chromodynamics
NICA,Nuclotron-based Ion Collider fAcility
FAIR,Facility for Antiproton and Ion Research
QGP,Quark–gluon plasma
RHIC,Relativistic Heavy Ion Collider
NM,Normal matter
PC,Pion condensation
NICER,Neutron star Interior Composition Explorer
NJL,Nambu–Jona-Lasinio
PNJL,Polyakov-loop-extended Nambu–Jona–Lasinio
CMF,Chiral Mean Field
LIGO,Laser Interferometer Gravitational-Wave Observatory
\end{filecontents*}

\setcounter{section}{0} 
\section{Introduction to Critical Points}
In thermodynamics, a critical point {refers} to the end point of a phase equilibrium curve. The presence of a critical point was first identified by Charles Cagniard de la Tour in 1822 ~\cite{1822annales}. He observed that, above a critical temperature, there was a continuous transition of elements in a gun barrel from liquid to vapour. A couple of decades later, Faraday (1845) studied the liquefying of gases, and  { Mendeleev (1860) ~\cite{mendeleev1860chastichnoe}} measured a vanishing point of liquid--vapour 
 surface tension naming it ``Absolute boiling temperature''~\cite{faraday:1843liquefaction}. Finally, in 1869, Andrews studied many different substances, establishing the continuity of vapour--liquid phases and creating the name and modern concept of ``critical point''~\cite{andrews:1869}.
In theoretical research,  { van der Waals(1873)~\cite{van1873over} }contributed by providing an equation of state with a critical point for the liquid--gas transition. {Smoluchowski~\cite{v1908molekular} and Einstein ~\cite{einstein1910theorie}  (1908, 1910)} explained critical opalescence, the {phenomenon} by which a liquid at the critical point becomes opaque to a collimated beam of laser light, which results from fluctuations becoming comparable in size with the wavelength of light, and causing scattering. These led to Landau's classical theory of critical phenomena, Fisher, Kadanoff, and Wilson scaling, and, eventually, to a full {fluctuation-based formulation} of renormalization group theory~\cite{Stephanov:1998dy}.

The question arises whether there exists a critical point associated with the phase expected to be {found} at high-energy density, namely chirally symmetric {matter composed of} deconfined quarks and gluons, similar to {the critical point of} liquid-vapour {transitions}. Quantum Chromodynamics describes the fundamental constituents of matter, quarks and gluons, as being almost massless, however, hadrons are massive. The mass of hadrons, which is associated with the breaking of chiral symmetry, is the source of most visible mass {in} the universe. The constituents of QCD carry color charge and experience the strong force, although they are normally confined within the hadrons. High-energy collisions deconfine those charges, producing the quark-gluon plasma. Finding the QCD critical point experimentally in heavy-ion experiments is a formidable task currently being pursued at the Relativistic Heavy Ion Collider (RHIC) \cite{bzdak2020mapping,an2021best}. Additional studies will be possible when dedicated facilities enter into operation, such as {Nuclotron-based Ion Collider fAcility (NICA) \cite{Senger:2017nvf} and Facility for Antiproton and Ion Research (FAIR)} \cite{Kekelidze:2017tgp}.

At zero baryon chemical potential, $\mu_B=0$, baryons and antibaryons or quarks and antiquarks are in chemical equilibrium (with respect to the strong, not the weak force) with one another. In this case, at finite temperature, it is possible to determine the QCD pressure and additional thermodynamic quantities numerically with controllable accuracy on the lattice. Calculations at finite quark masses, in which case the chiral symmetry of QCD is not exact, do not exhibit the presence of a critical point (at $\mu_B=0$). However, notable changes in properties of matter with increasing temperature appear. The significant rise in pressure with increasing temperature indicates the change in phase from a hadron gas to a quark-gluon plasma. However, this transition is not sharp. In other words, there is no point at which those phases coexist, despite one phase continuously transforming into another, which is referred to as a crossover \cite{Aoki:2006we}. 

The approximate location of a critical point when $\mu_B \neq 0$, {may be found through} extrapolations performed by calculating coefficients of a Taylor expansion around $\mu_B = 0$ and using assumptions from analyticity and convergence \cite{PhysRevD.71.114014,fodor2019trying}. Complementary approaches using holographic gauge/gravity correspondence to map the fluctuations of baryon charge in the dense quark-gluon liquid onto a numerically tractable gravitational problem involving the charge fluctuations of holographic black holes have been able to predict the position of the QCD critical point at {$T = 89$ MeV , {$\mu_B = 724$}} MeV \cite{Critelli:2017oub}.  For other theoretical models that also predict the existence and position of the QCD critical point see References~\cite{Halasz:1998qr,Stephanov:2004wx,ayala2021collision}. 

Alternatively to $\mu_B$, QCD phase diagrams are also studied in the temperature-isospin chemical potential, $\mu_I$, plane. The later is advantageous due to the fact that lattice QCD results are not afflicted by the sign problem at finite $\mu_I$, as long as $\mu_B=0$. When $\mu_B \neq 0$ or, {equivalently, when} there is a difference in the number of particles and anti-particles in the system, the first-principle methods such as non-perturbative lattice QCD simulations cannot be implemented \cite{Karsch:2001cy,Muroya:2003qs,Bedaque:2017epw}. Finite $\mu_I$, zero $\mu_B$ calculations have recently received a lot of attention due to their relevance to heavy-ion collisions~\cite{Li_1998}, compact stars~\cite{Migdal:1990vm,Steiner_2005,PhysRevD.98.094510} and even to the evolution of the early Universe~\cite{Schwarz_2009}. These calculations can be used to verify if the predictions of effective models of QCD are reliable when {compared} with first principle calculations~\cite{Cotter:2012mb,Braguta:2016cpw,Bali:2012zg,Bali:2011qj,Bali:2014kia,Braguta:2015zta,Braguta:2015owi}. In the case of QCD at finite isospin density, the first lattice simulations {of References~\cite{PhysRevD.66.034505,PhysRevD.66.014508} were performed using} unphysical pion masses and/or unphysical flavor content. The results from these first simulations were in qualitative agreement with the different approaches to QCD~\cite{PhysRevLett.86.592,Son:2000by,PhysRevD.99.096011,Carignano:2016lxe,PhysRevD.93.094502,Cohen:2015soa,PhysRevD.79.014021,PhysRevD.67.074034,PhysRevD.71.094001,PhysRevD.64.016003,Mannarelli:2019hgn,PhysRevD.95.105010,PhysRevD.98.054030,Khunjua:2018jmn,PhysRevD.88.056013,PhysRevD.82.056006,PhysRevD.79.034032,Andersen_2009,PhysRevD.75.096004,Ebert:2005wr,Ebert_2006,PhysRevD.74.036005,PhysRevD.71.116001,He:2005sp,PhysRevD.69.096004,Toublan:2003tt,Frank:2003ve,PhysRevD.75.094015,PhysRevC.89.064905,Andersen:2015eoa,PhysRevD.98.074016,Stiele:2013pma,PhysRevD.88.074006,Kamikado:2012bt}.  More recently, lattice QCD results for finite isospin density were implemented using an improved lattice action with staggered fermions at physical quark and pion masses \cite{PhysRevD.97.054514,Brandt:2017zck,Brandt:2016zdy,Brandt:2018wkp}, their predictions being in very good agreement with the results obtained from updated chiral perturbation theory~\mbox{\cite{Adhikari:2019zaj,Adhikari:2019mlf,Adhikari:2019mdk}} and {Nambu–Jona-Lasinio (NJL)} models~\cite{Avancini:2019ego,Lopes:2021tro,Lu:2019diy,Wu:2020knd}.

For $\mu_B=0$, in the regime of small temperatures and small $\mu_I$, the system is said to be composed of so-called normal matter, NM, without pion condensation. Increasing only the temperature, the system enters a phase in which chiral symmetry is partially restored and quarks are deconfined, that is the expected quark-gluon plasma phase. On the other hand, for low temperature and increasing $\mu_I$, the system reaches the onset of the pion condensation, PC, phase through a second-order phase transition. The PC onset takes place at $\mu_I=m_{\pi}$. For $\mu_I < m_\pi$ there is a smooth crossover related to the chiral symmetry restoration and deconfinement transitions, and the pion condensate vanishes for all temperatures. For $\mu_I > m_\pi$, one can find a critical temperature at which there is a second order phase transition from the PC to the QGP phase. In this case, there is a lattice QCD prediction for a pseudo-triple point, where the NM, quark-gluon plasma, and PC phases meet \cite{PhysRevD.97.054514,Brandt:2018wkp}. There are also other calculations, such as a recent work using the non-local {Polyakov-loop-extended Nambu–Jona–Lasinio (PNJL) model} \cite{carlomagno2021isospin}, which is in excellent agreement with lattice QCD calculations for the QCD phase diagram at finite $\mu_I$.

In this work, we combine both approaches discussed above and study three-dimensional QCD phase diagrams as a function of temperature, baryon chemical potential, and either charge fraction, $Y_Q$, or isospin fraction, $Y_I$. The relation among the two charges is trivial only when net strangeness is zero. A fixed fraction is achieved by self-consistently determining the respective (charge or isospin) chemical potential. In this work, we consider two particular cases, one in which strangeness is constrained to be zero (same amount of strange and anti-strange particles, achieved by self-consistently determining the respective strange chemical potential, $\mu_S$), and one in which strangeness is free to vary (in which case $\mu_S=0$).
We focus, in particular, on how the end point of the first-order phase transition line changes for particular cases associated with different astrophysical and laboratory scenarios.

\section{Our Formalism}

In order to study the effects of different conditions on the critical point, we first describe the formalism that allows us to reproduce a deconfinement first-order phase transition that eventually (at large temperatures) becomes a crossover. {This phase transition is necessarily tied to a chiral-symmetry restoration phase transition (independently of the transition order), as they are connected through the effective masses of hadrons and quarks, in which the respective order parameters for both kinds of phase transition appear (see Equations~(67) in Reference~\cite{Dexheimer:2009hi}).} It is very important to note that such a formalism must contain the degrees of freedom expected to be found on both sides of the coexistence line in order to fully determine how conditions such as strangeness affect these degrees of freedom and, ultimately, moves the coexistence line and critical end point to a different part of the QCD phase diagram. {We satisfy this requirement by making use of the} Chiral Mean Field (CMF) model {which} is an effective relativistic mean-field model that describes baryons, quarks, and leptons within the same formalism \cite{Dexheimer:2009hi}. It is based on a non-linear realization of the sigma model \cite{Papazoglou:1998vr} {within the mean-field approximation}.

{The model} has been shown to give a good description of the dense matter inside proto-neutron and neutron stars \cite{Roark:2018boj,Dexheimer:2020rlp}. {This includes reproducing i) $2$ solar masses neutron stars, or even more massive ones when higher-order vector interactions are used \cite{Dexheimer:2020rlp}, ii) radii that agree with { The Laser Interferometer Gravitational-Wave Observatory (LIGO)}  results \cite{LIGOScientific:2018cki} provided throught the use of spectral equation of state, or both methods (throught the use of spectral equation of state and universal relations \cite{Yagi:2015pkc,Yagi:2016qmr}) when vector isovector couplings are included (all $90\%$ confidence) \cite{Dexheimer:2018dhb}, and iii) { The Neutron star Interior Composition Explorer (NICER)} results for radii ($90\%$ confidence) \cite{Riley:2019yda,Miller:2019cac,Riley:2021pdl,Miller:2021qha}.
We have also analyzed  the differences that arise from the relaxation of the assumption of local conserved quantities, such as electric charge and lepton fraction, to global quantities (the latter for the first time)~\cite{Roark:2018uls}. We found that, although such relaxation widens the phase transition as a function of baryon chemical potential or free energy (less for the fixed lepton fraction case), there is no change in the position of the critical point. This is a result of the dominance of thermal effects around the deconfinement critical point  which is not the case for the nuclear liquid--gas phase transition, as discussed in detail in Reference~\cite{Hempel:2013tfa}. Since in this work we focus on the discussion of critical points, we assume that the surface tension of quark matter is very large (of the order of hundreds of MeV), in which case mixtures of phases do not appear and ours constraints related to strangeness, charge, and isospin can be carried out locally. More complicated cases will be studied in the future.}

At low densities, the model was fit to nuclear saturation properties at zero temperature and lattice QCD data at large temperatures. {The latter includes a first order phase transition at $T = 270$
MeV and a pressure functional P(T) similar to References~\cite{Roessner:2006xn,Ratti:2005jh} at vanishing baryon chemical potential for pure gauge, a crossover at vanishing baryon chemical potential with a transition temperature of $171$ MeV (determined as the peak of the change of the chiral condensate and the parameter for deconfinement), and the location of the critical endpoint (at $\mu_{B_{cr}}= 354$ MeV, $T_{cr} = 167$ MeV for symmetric matter) in accordance with Reference~\cite{Fodor:2004nz}.}
{In the past, this formalism was used to simulate particle multiplicities, rapidity distributions, and flow in heavy-ion collisions \cite{Steinheimer:2009nn} and} the differences among stellar matter vs. matter produced in heavy-ion collisions in the vicinity of the deconfinement coexistence line, drawing analogies with the nuclear liquid--gas phase  diagram \cite{Hempel:2013tfa}. More recently, we used this formalism to construct three-dimensional QCD phase diagrams {up to a} temperature {$T = 160$ MeV, the other two axes being the} chemical potential ($\mu_B$ or $\tilde{\mu}$), and either the charge or isospin fractions \cite{Aryal:2020ocm}. In this work, we go further by extending coexistence lines all the way to the respective critical end points and discussing the effects of different conditions on the relative position of the critical end points.
In the following, we summarize the different {situations under consideration. We begin by defining} a few relevant quantities. 

The free energy per baryon under fixed charge and strangeness is defined as \cite{PhysRevD.80.125014}
\begin{eqnarray}
 \tilde{\mu} = \mu_B + Y_Q\mu_Q +Y_S\mu_S\ ,
 \label{eqn:muhat1}
\end{eqnarray}
{where $\mu_Q$, $\mu_I$, and $\mu_S$ are the charge, isospin, and strange chemical potentials, respectively. Alternatively,} under fixed isospin and strangeness \cite{Aryal:2020ocm}
\begin{eqnarray}
&\widetilde{\mu} &= \mu_B' + Y_I\mu_I + Y_S\mu_S' ,
\label{eqn:mub_mus}
\end{eqnarray}
 where, under the new constraints, the chemical potentials transform as $\mu_I=\mu_Q$, $\mu_B' = \mu_B+ \frac{1}{2} \mu_Q$, and $\mu_S' = \mu_S - \frac{1}{2} \mu_Q$. These definitions reproduce the same equilibrium equations described in Appendix A of Reference~\cite{Aryal:2020ocm}, independently of the chosen constraints. The free energy per baryon $\tilde{\mu}$ is the quantity which is the same in both phases in the case of a first-order phase transition (both sides of the coexistence line). The charge fraction, $Y_Q = \frac{Q}{B}$, isospin fraction, $Y_I =\frac{I}{B}$, and strangeness fraction, $Y_S=\frac{S}{B}$, are defined as the {baryonic (hadrons + quarks) electric charge}, isospin, and strangeness over the total baryon number. The relation between charge and isospin fractions depends on the strangeness following Reference~\cite{Aryal:2020ocm} is
\begin{eqnarray}
Y_I =Y_Q - \frac{1}{2} + \frac{1}{2}Y_S\ . 
\end{eqnarray}
\label{eqn4}
\vspace{-15pt}

\section{Results}

\begin{figure}[t!]
\begin{subfigure}[t]{0.36\textwidth}
\includegraphics[width=\textwidth]{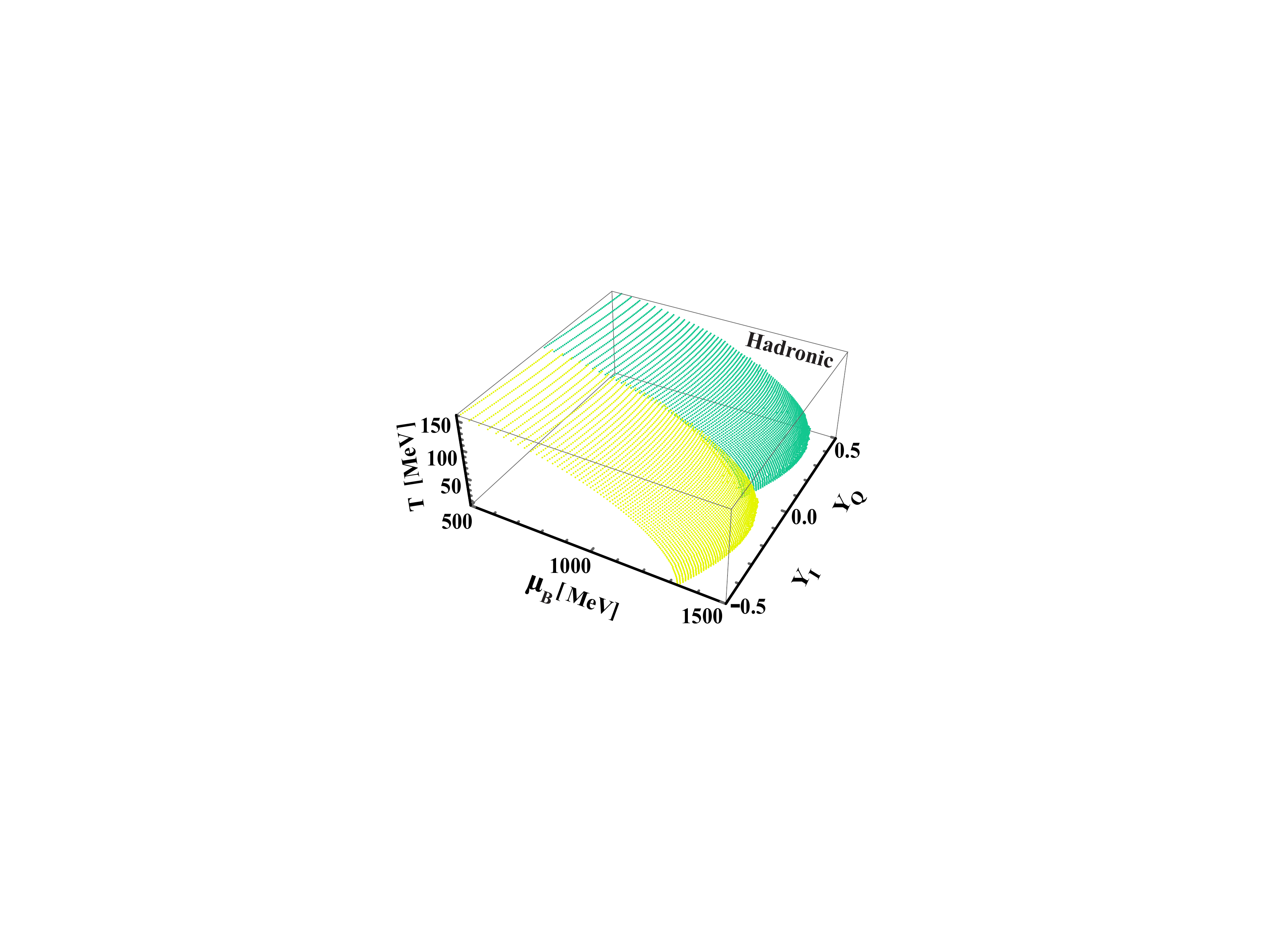}
  \end{subfigure}
 \begin{subfigure}[t]{0.36\textwidth} 
 \includegraphics[width=\textwidth]{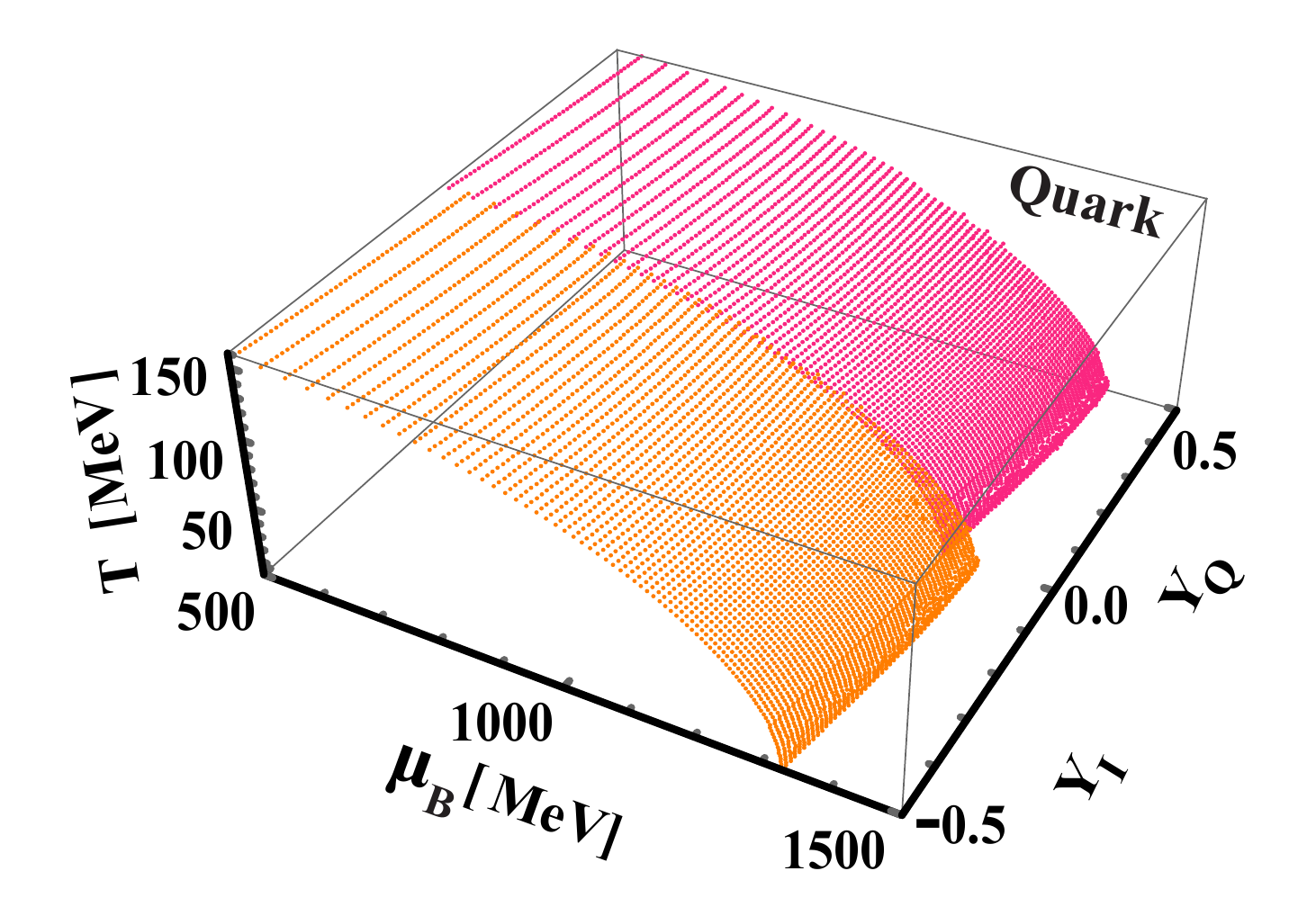}
  \end{subfigure}
\begin{subfigure}[t]{0.36\textwidth} 
\includegraphics[width=\textwidth]{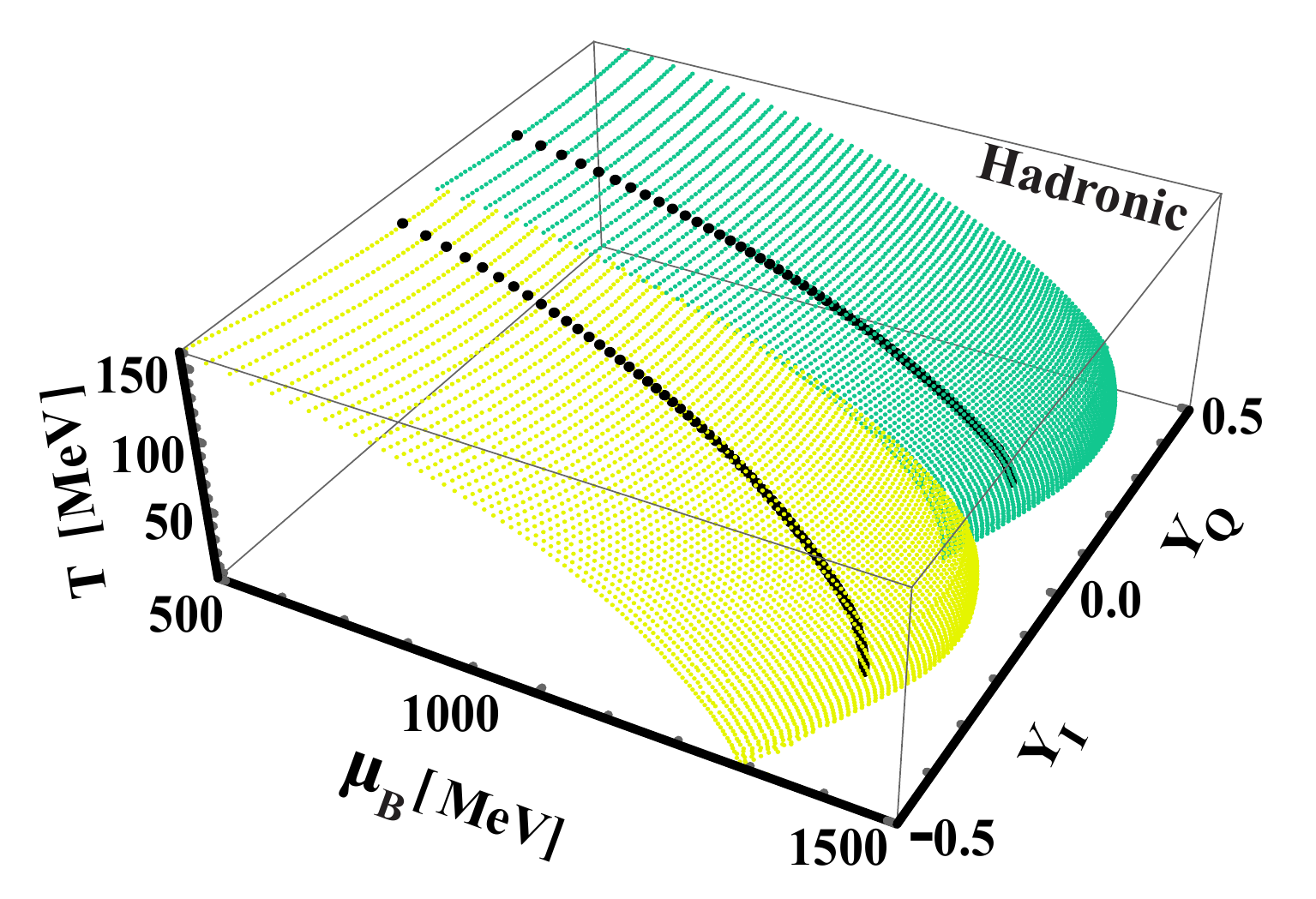}
\end{subfigure}
  \begin{subfigure}[t]{0.36\textwidth}  
\includegraphics[width=\textwidth]{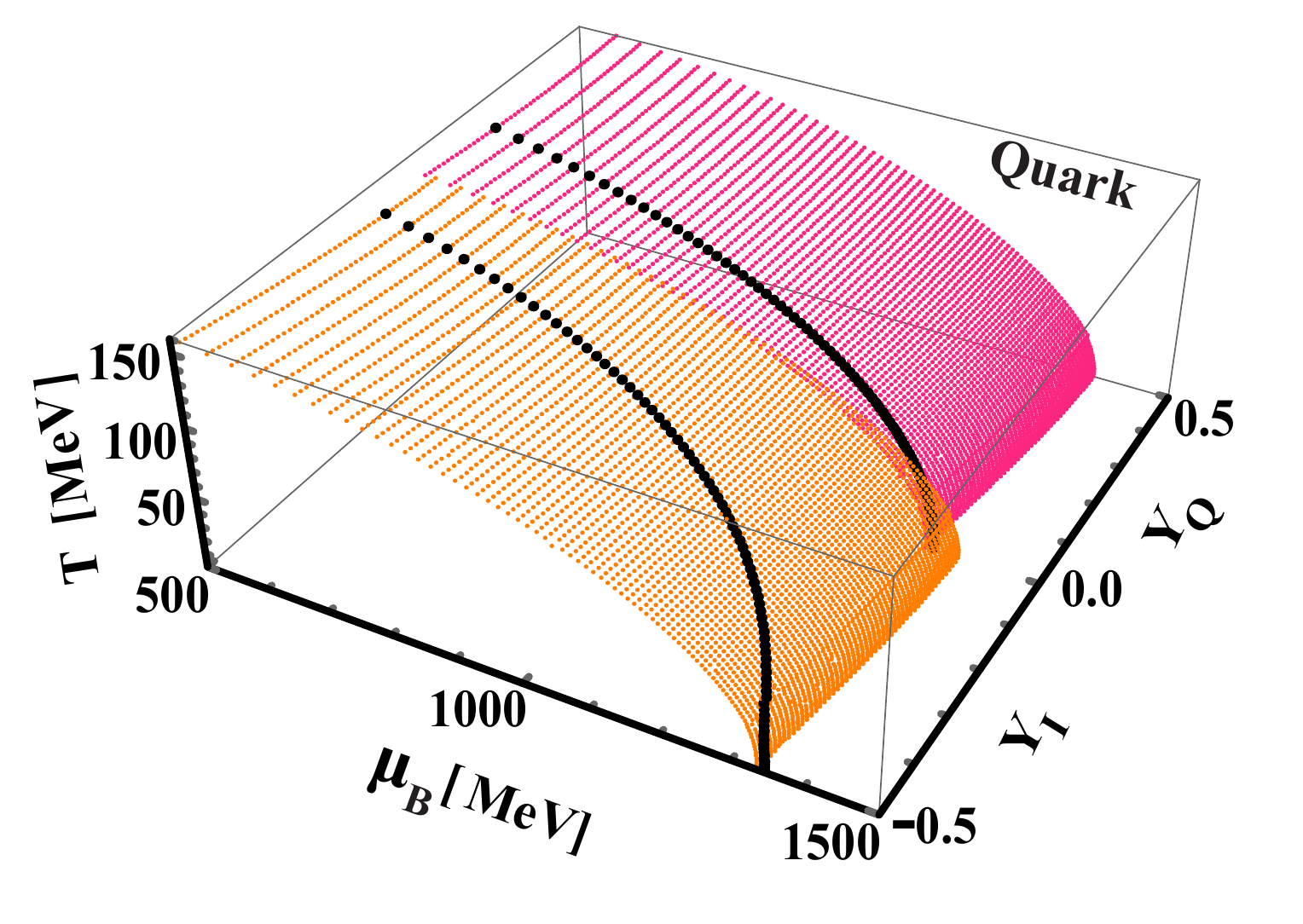}
\end{subfigure}
\caption{l{Three-dimensional} QCD phase diagrams showing temperature, baryon chemical potential, and either charge fraction (dark green/pink regions) or isospin fraction (light green/orange regions regions). The left panels show the hadronic side of the coexistence line, while the right panels show the quark side. The top panels show matter with zero net strangeness ($Y_S=0$), while the bottom panels show matter with net strangeness ($Y_S\neq0$). The black lines on the bottom panels show charge neutral, strange matter in chemical equilibrium with leptons.} 
\label{phase3D}
\end{figure}

The different conditions we discuss in this work relate to systems created in very short-lived heavy-ion collisions, where there is no sufficient time for significant net strangeness to be produced, $Y_S=0$, and astrophysical conditions, where net strangeness is expected to be produced, $Y_S\neq0$. While matter generated in heavy-ion collisions and supernova explosions is close to being isospin symmetric, matter in neutron stars and their mergers \cite{Most:2019onn} is far from that, possessing very low charge fraction. Some simple cases that approximate these conditions are:
\begin{itemize}
    \item $Y_Q=0$ and $Y_S=0$ (equivalent to $Y_I=-0.5$ and $Y_S=0$)\ ;
    \item $Y_Q=0.5$ and $Y_S=0$ (equivalent to $Y_I=0$ and $Y_S=0$)\ ;
    \item $Y_Q=0$ and $Y_S\neq0$\ ;    
    \item $Y_I=-0.5$ and $Y_S\neq0$\ ;
    \item $Y_Q=0.5$ and $Y_S\neq0$\ ;     \item $Y_I=0$ and $Y_S\neq0$\ ;
    \item $Y_Q=-Y_{\rm{lepton}}$ and $Y_S\neq0$\ ,
\end{itemize}
where in the last case electrons and muons were introduced in chemical equilibrium with the baryons and quarks, $\mu_e=\mu_\mu=-\mu_Q$. {Note that, despite first appearances, conditions 3-4 and 5-6 are \textit{not} equivalent. As already mentioned in Section 1, this is because the relation between $Y_I$ and $Y_Q$ is a trivial shift only when net strangeness is zero, but not otherwise. See Reference~\cite{Alford:2021ogv} for a complete discussion of the chemically equilibrated case at finite temperature.}
Figure~\ref{phase3D} shows three-dimensional deconfinement first-order phase-transition regions calculated finding a discontinuity in the order parameter for deconfinement. The left panels show values for the baryon chemical potential $\mu_B$ on the hadronic side, while the right panels show values for $\mu_B$ on the quark side of the coexistence region. Their difference stems from the fact that we are plotting them as a function of baryon chemical potential (not free energy per baryon) and that $\mu_Q$ or $\mu_I$ is different in each phase. The different colors show calculations performed by varying the charge fraction from $0\to0.5$ (shown as a function of $Y_Q$) and the isospin fraction from $-0.5\to0$ (shown as a function of $Y_I$). The top panels were calculated fixing the net strangeness to zero and, as a consequence, are a perfect $0.5$ shift in the fraction axis (see Equation~(3)). The bottom panels had no strangeness constraint and, therefore, present a finite net strangeness. {As a result, the two pairs of surfaces are not only shifted but also \textit{deformed} relative to one another.} This {leads to more pronounced} differences (from the top panels) at large temperatures, where strangeness is abundant. The bottom panels also show a chemically equilibrated line which, at low temperatures, follows a much lower value of charge or isospin fraction in the quark phase. 

\begin{figure}[t!]
\includegraphics[width=12cm]{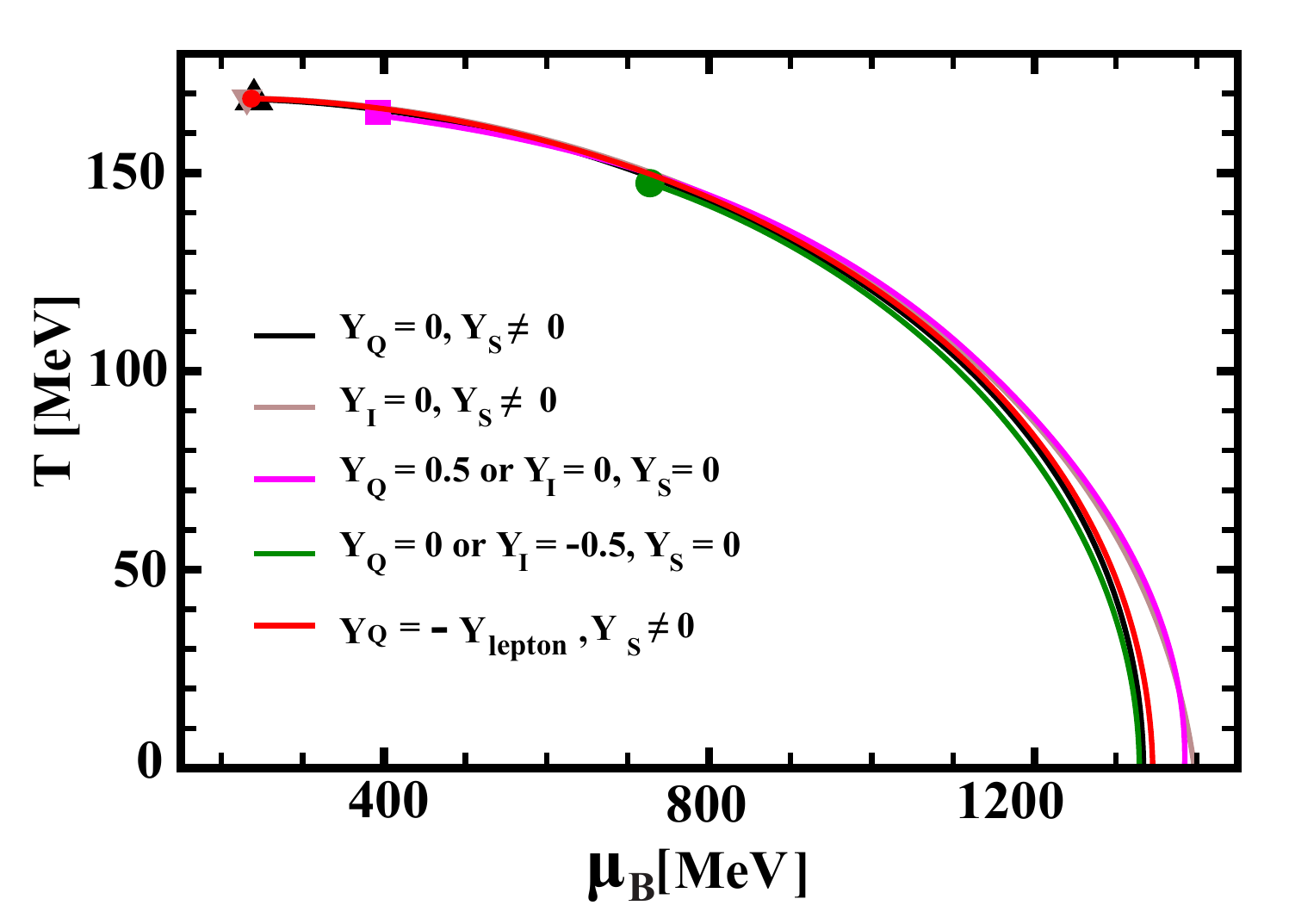}
\caption{{Temperature} versus baryon chemical potential for several cases with different fixed charge or isospin fractions with and without net strangeness, plus the particular case of chemical equilibrium with leptons. In all cases the baryon chemical potential equals the free energy per baryon.}
\label{fig:T_vs_mub}
\end{figure}

At the largest temperature we show in Figure~\ref{phase3D}, $160$ MeV, it is easy to see that the phases are very similar, but no further analysis was made to find the exact position of the critical end point. In some cases, our region went into the crossover phase, marking the steepest change in order parameters (peak of susceptibility) instead of the position of the first-order phase transition. 
{The critical end point occurs when the pressure (grand-potential density) as a function of the deconfinement order parameter, exhibits a single maximum (minimum) instead of two. For a more focused exploration of the exact position of our critical end point, we present in Figure~\ref{fig:T_vs_mub} two-dimensional phase diagrams for deconfinement for the five of the seven cases described in the previous section. These are the cases in which} the free energy per baryon equals the baryon chemical potential. From Equation~(1) {it} is quite obvious that this happens for $Y_Q=0$. It also happens when $Y_I=0$, meaning that $\mu_Q=0$. This is the case no matter if the strangeness is being constrained or not. In the chemical equilibrium case, the total charge ($Y_Q+Y_{\rm{lepton}}$) is zero, implying once more   $\tilde{\mu} = \mu_B$, but we only investigate this for the case with net strangeness, as this is the case associated with chemically-equilibrated stars.

Analyzing Figure~\ref{fig:T_vs_mub} in more detail, one can see that both lower charge or isospin fraction and the constraint of zero net strangeness weaken the phase transition, shifting the critical end point to a lower temperature (and, as a consequence, to a larger $\mu_B$). When combined, these features cause the green curve to shrink and the green dot to be found more than $21$ MeV below the beige inverted triangle, which represents the ``opposite'' case ($Y_I=0$ and $Y_S\neq0$). A lower critical temperature means that the first order transition between the phases becomes weaker faster with the increase of temperature { compared to the other cases}. This can be seen in the size of the discontinuity in energy density or other derivatives of the grand potential across the transition and it is related to how different or similar the two phases are. The coexistence line for the {chemically-equilibrated} case is clearly more similar to the $Y_Q=0$ cases at low temperatures (when $\mu_Q$ is more negative, see Figure~2 of \cite{Dexheimer:2020okt}), and becomes slowly more similar to the $Y_S\neq0$ at large temperatures, when strangeness is abundant. 

\begin{figure}[t!]
\includegraphics[width=12 cm]{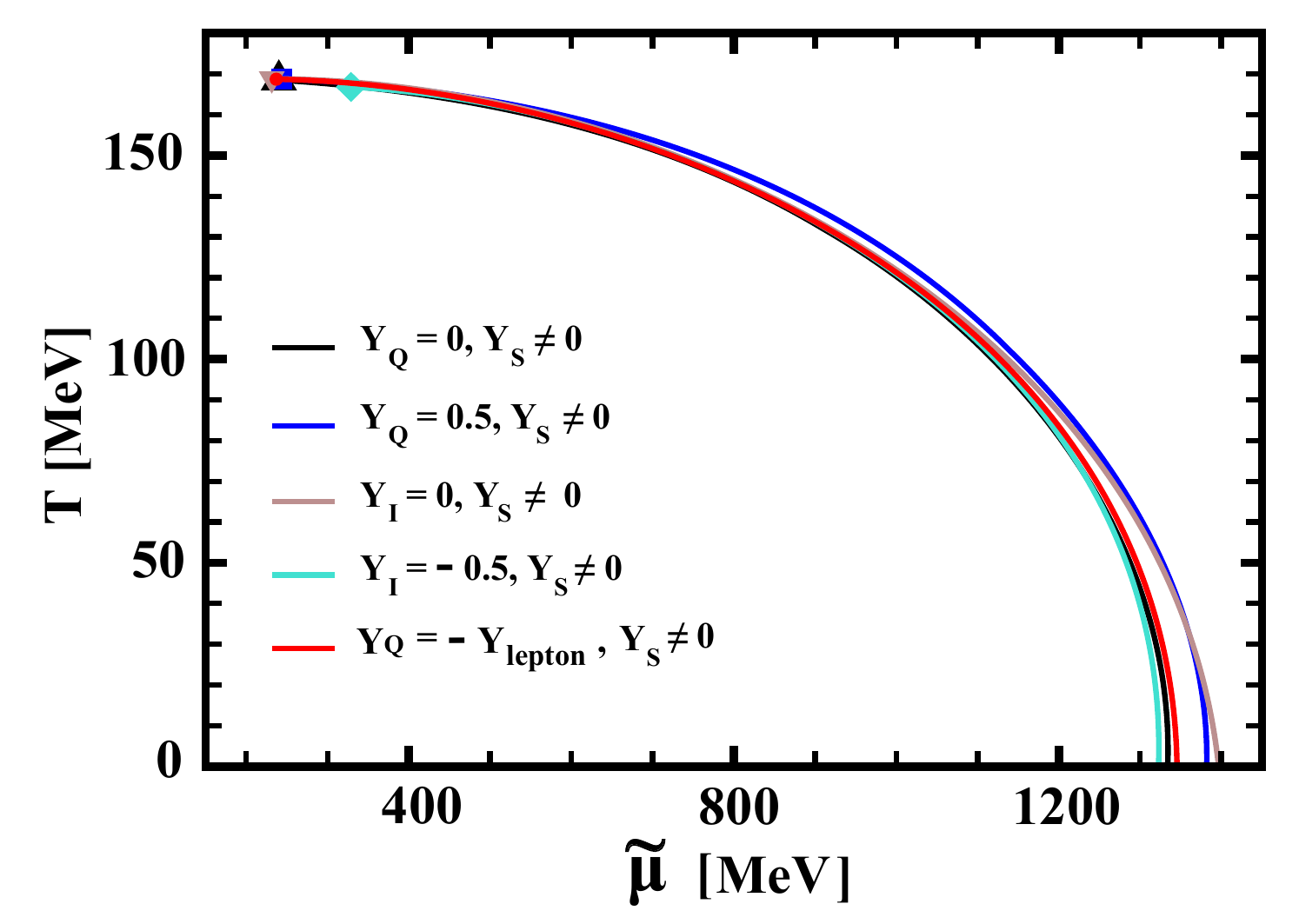}
\caption{{Temperature} versus free energy per baryon for several cases with different fixed charge or isospin fractions but always with net strangeness, plus the particular case of chemical equilibrium with leptons. Strangeness is not constrained.} 
\label{fig:T_vs_muhat}
\end{figure}

Numerical values for the critical end point and for the $T=0$ portion of the phase diagram are given in Table~\ref{tab:my-table}. For some of the cases, the baryon chemical potential is different in each phase which, again, traces back to Equation~(3). The cases in which this happens, namely $Y_Q=0.5$ and $Y_I=-0.5$ in the presence of net strangeness, are shown in Figure~\ref{fig:T_vs_muhat}, together with the other cases with strangeness for comparison. Here the critical end point is only distinguishable for the $Y_I=-0.5$ case, in which the critical temperature is a bit lower and the free energy per baryon larger than the other cases. It was shown in Reference~\cite{liu2021isospin}, using the NJL and the {PNJL} models to study instabilities in quark matter related to the liquid--gas phase transition, that large isospin asymmetries (more negative $Y_I$) and net strangeness weaken the first-order phase transition, making it easier for it to become a crossover. This is in partial agreement with what we found in this work (we found the opposite with respect to strangeness). This difference is related to the fact that in our case the critical end point is strongly affected by the presence of hadrons.

Concerning the general position of the lines in Figure~\ref{fig:T_vs_muhat}, as discussed in detail in Reference~\cite{Aryal:2020ocm} for the specific case of zero and large temperatures, larger charge fractions or smaller isospin fractions (both meaning more balanced amount of protons and neutrons or up and down quarks when not considering strangeness) push the phase transition to larger baryon chemical potentials at a given temperature. This has to do with how much more this change softens hadronic matter, so the quark phase becomes more energetically favored at lower energy densities, which corresponds to larger baryon chemical potentials. The coexistence line for the {chemically-equilibrated} case is clearly more similar to the $Y_Q=0$ and $Y_I=-0.5$ cases at low temperatures.


\begin{specialtable}[t!]
\small
\caption{Summary of baryon chemical potential, free energy per baryon, and temperature at the critical end point and the baryon chemical potential (in either side of the coexistence line) and free energy per baryon at zero temperature under differing conditions of strangeness, fixed charge/isospin fraction or in chemical equilibrium with leptons. All units are in  MeV.}
\label{tab:my-table}
\setlength{\tabcolsep}{4.5mm}
\begin{tabular}{ccccccc}

\noalign{\hrule height 1.0pt} 
\rowcolor{black!30}\textbf{{Cases}}  

& \boldmath{${\mu_B}_{cr}$} & \boldmath{${\tilde{\mu}}_{cr}$} & \boldmath{${T}_{cr} $} & \boldmath{${\mu_B}^Q_{0}$} & \boldmath{${\mu_B}^H_{0}$}& \boldmath{${\tilde{\mu}}_{0}$} \\
\hline
$Y_Q = 0,   Y_s \neq 0$ & $240.1$ & $240.1$ &$168.5$ & $1334.0$& $1333.0$& $1334.0$\\ 
\rowcolor{black!15}$Y_I = 0, Y_s \neq 0$ & $231.4$ & $231.4$ &$168.8$ &$1396.7$ & $1396.8$  & $1396.7$ \\
$Y_Q = 0.5, Y_s = 0$ & $392.7$  & $392.7$ &$165.2$ &$1382.0$ & $1381.0$ & $1382.0$\\
\rowcolor{black!15}$Y_Q = 0,   Y_s = 0$ & $727.1$ & $727.1$& $147.4$ &$1329.0$ & $1328.0$ & $1329.0$\\
$Y_Q = 0.5, Y_s \neq 0$  & $227.6$ & $244.0$ &$168.8$ &$1382.0$ & $1381.0$ & $1382.0$\\
\rowcolor{black!15}$Y_I = -0.5, Y_s \neq 0$ & $295.9$  & $328.9$ & $166.8$ & $1322.8$ & $1314.3$ & $1322.8$ \\
$Y_Q=-Y_{\rm{lep}}$, $Y_s \neq 0 $  & $236.9$  & $236.9$ & $168.7$ & $1345.5$ & $1345.4$& $1345.5$\\
\bottomrule
\end{tabular}

\end{specialtable}

\section{Conclusions}

After an introduction to the topic of critical points, we discussed qualitative results for critical end points reproduced by the { CMF model}. In order to do so, we first built three-dimensional phase diagrams showing the deconfinement coexistence region as a function of temperature, baryon chemical potential, and either charge fraction ,$Y_Q$, or isospin fraction, $Y_I$. We also analyzed different scenarios concerning zero or unconstrained net strangeness fraction, $Y_S$, {and} presented results for the particular case of weak chemical equilibrium with leptons.

We then focused on different specific cases for combinations of $Y_Q=0$, $Y_Q=0.5$, $Y_I=0$, $Y_I=-0.5$, weak chemical equilibrium, $Y_S=0$, and $Y_S\neq0$. The isospin symmetric and the large charge fraction cases present a later (in both baryon chemical potential and free energy per baryon) phase transition at low temperatures. These cases correspond to matter created in heavy-ion collisions or in supernova explosions. At large temperatures, all curves are close, but some extend to significantly lower baryon chemical potentials and free energies. The ones that {do not}, and therefore present a distinguished critical end point, are the non-strange cases, once more corresponding to matter created in heavy-ion collisions. This feature had not been investigated before within the CMF model.

For astrophysical conditions, corresponding to larger net strangeness content and negative isospin fraction (or low charge fraction), this means an earlier confinement phase transition (lower value of baryon chemical potential and free energy per baryon) and a still strong transition that extends high in temperature. This is an encouraging result, as it has been shown that strong first-order phase transitions can leave distinguishable signals in the post merger part of gravitational wave signals from neutron star mergers~\cite{Most:2018eaw}.


\authorcontributions{{All authors contributed significantly to this work}. All authors have read and agreed to the published version of the manuscript.} 
%
%
%

\funding{Support comes from ``PHAROS'' COST Action CA16214 and the National Science Foundation under grant
PHY-1748621. This work was partially supported by Conselho Nacional de Desenvolvimento Cient\'ifico 
e Tecno\-l\'o\-gico  (CNPq), Grant No. 309598/2020-6 (R.L.S.F.); 
Funda\c{c}\~ao de Amparo \`a Pesquisa do Estado do Rio 
Grande do Sul (FAPERGS), Grants Nos. 19/2551- 0000690-0 and 19/2551-0001948-3 (R.L.S.F.). {C.C. acknowledges support from the European Union’s Horizon 2020 research and innovation programme under the Marie Sk\l{}odowska-Curie grant agreement No. 754496 (H2020-MSCA-COFUND-2016 FELLINI).}}

\institutionalreview{{text.}}

\informedconsent{{text.}}

\dataavailability{{text.}}

\conflictsofinterest{The authors declare no conflict of interest.}

\section*{List of Acronyms}

\begin{itemize}
\DTLforeach*{acronyms}{\thisAcronym=Acronym,\thisDesc=Description}%
  {\item \textbf{\thisAcronym} \thisDesc}%
    \end{itemize}

\end{paracol}
\reftitle{References}

\end{document}